\documentclass[twocolumn,showpacs,preprintnumbers,amsmath,amssymb,nofootnoteinbib]{revtex4}

\usepackage{graphicx,subfigure}% Include figure files
\usepackage{dcolumn}% Align table columns on decimal point
\usepackage{bm}% bold math
\usepackage{psfrag}
\usepackage{color}
\usepackage{slashed}

\newcommand{\be}{\begin{equation}}
\newcommand{\ee}{\end{equation}}

\def\bea{\begin{eqnarray}}
\def\eea{\end{eqnarray}}

\begin{document}

\bibliographystyle{apsrev}

\title{A model solving the PVLAS-CAST puzzle}

\author{Javier Redondo}

\affiliation{Grup de F{\'\i}sica Te{\`o}rica and Institut de F{\'\i}sica d'Altes
Energies\\Universitat Aut{\`o}noma de Barcelona\\
08193 Bellaterra, Barcelona, Spain}

\begin{abstract}
Axion physics has received a boost with the recent claim of the PVLAS collaboration. Their results
can be interpreted as due to a new axionlike particle but the CAST collaboration has not found
trace of it. While the axionlike particle interpretation of the PVLAS signal is going to be probed
by dedicated laboratory experiments, it is mandatory to find either alternatives to it or models in
which the astrophysical bounds could be evaded. In this communication one of such models is
presented. The new physics involved appears at a low energy scale $<$ eV.
\end{abstract}

% PACS, the Physics and Astronomy

\maketitle

\section{The PVLAS $\&$ CAST puzzle}
The PVLAS collaboration has recently claimed a rotation of the polarization of laser light
propagating through a transverse magnetic field \cite{Zava06}. This can be interpreted as the
oscillation of photons into very light axionlike particles $\phi$ (ALPs hereafter) through an
interaction like
\begin{equation}
\frac{1}{4M}\, F^{\mu\nu}F_{\mu\nu}\, \phi  \hspace{.7cm} \rm or \hspace{.7cm} \frac{1}{8M}\,
\epsilon_{\mu\nu\rho\sigma}F^{\mu\nu} {F}^{\rho\sigma}\, \phi \label{ALP_coupling}
\end{equation}
depending on the sign of the ALP parity. Here $F_{\mu\nu}$ is the photon field strength. The PVLAS
rotation plus the previous bounds of the BRFT experiment \cite{Came93} imply \cite{Ring06a}
\begin{equation}
2 \times 10^5  < \frac{M}{\rm GeV} < 6\times 10^5\ ;\ 1  < \frac{m_\phi}{{\rm meV}} < 1.5
\end{equation}
for the dimensionful coupling $M$ and the ALP mass $m_\phi$.

However, such a not-so-weakly-interacting axionlike particle would be produced by Primakoff effect
in the interior of stars releasing a huge amount of energy very fast and thus cooling the star in a
time scale which is hardly compatible with observations. Qualitatively it is found that couplings
satisfying
\begin{equation}
M > 1.7  \times 10^{10}\  {\rm GeV} \label{M_star}
\end{equation}
are harmless for evolutionary time scales of HB stars in globular clusters \cite{Raff96}.

Moreover, the CAST helioscope could detect axionlike particles coming from the Sun  by inverse
Primakoff effect in a strong magnetic field. No signal has been found, resulting in the bound
\cite{Ziou04}
\begin{equation}
M > 0.87 \times 10^{10}\ {\rm GeV}  \ \ \ . \label{M_CAST}
\end{equation}
Both constraints make strongly unlikely the ALP interpretation of PVLAS. However, we find both of
them based on the hypothesis that the ALP coupling (\ref{ALP_coupling}) holds equally well for two
very different environments: The high vacuum permeated by a magnetic field of PVLAS experiment and
the hot and dense plasmas of the interior of the Sun and HB stars.

Inspired by this ideas we have suggested in \cite{Mass05} that the value of $M$ could decrease with
the momentum transfer $q^2$ at which the interaction (\ref{ALP_coupling}) is probed. Note that in
PVLAS $\sqrt{q^2}\sim 10^{-6}$ eV while in the Primakoff conversion in stars $\sqrt{q^2}\sim$ keV.
This idea has been generalized in \cite{Jaec06b} considering also a dependence of environmental
parameters such as temperature, matter or energy density which are also very different in both
environments. Following very simple criteria we find that it is possible although extreme to evade
the astrophysical bounds (\ref{M_star}) and (\ref{M_CAST}) to reconcile the PVLAS ALP with the CAST
and astrophysical bounds. The conditions seem to be new physics at an accessible scale.

Many authors have followed these ideas in very different models. In \cite{Anto06} the ALP is
substituted by a spin 1 particle that decouples of light at the keV energies of stellar
environment. Also, in \cite{Moha06},  a model is proposed in which $1/M$ is proportional to the
vacuum expectation value of an additional scalar field which vanishes at stellar conditions.

However, these are not the only ideas suggested to solve the PVLAS-CAST puzzle. In
\cite{Mass05,Jain05} it is proposed that the ALP could interact strongly within stellar plasma thus
evading the CAST detection window and in \cite{Ferr06} the authors claim that the PVLAS signal is
due to purely QED effects. None of this models is completely satisfactory for the moment.

In this short communication I want to report a model \cite{Mass06a} which falls into the first
category. It consists in a UV competition of (\ref{ALP_coupling}) which contains new physics that
makes this ALP coupling much weaker in stellar environments. The proposed structure is a loop in
which a new fermion $f$ runs (See Fig.\ref{triangle}). The matching between this microphysical
picture and (\ref{ALP_coupling}) gives
\begin{equation}
\frac{1}{M} = \frac{\alpha}{\pi} \frac{Q_f^2}{v} \label{triangle}
\end{equation}
where $\alpha=e^2/4\pi$ is the fine structure constant and $Q_f$ is the electric charge of $f$
normalized to $e$. Here $v$ is a function of $m_f$ and $m_\phi$ if $\phi$ has a scalar (or
pseudoscalar) coupling with $f$ but can be an completely independent energy scale if $\phi$ is a
Goldstone boson\footnote{The PVLAS collaboration has also performed measurements of birefringence
of the magnetic field, which suggest that the coupling is of scalar type \cite{Gast06}}.
\begin{figure}[h]
  \vspace{2cm}
  \includegraphics{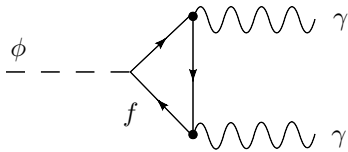}\vspace{0cm}
  \caption{\it
    Triangle diagram for the $\phi\gamma\gamma$ vertex.
   \label{fig1} }\vspace{0cm}
\end{figure}

As we need $v$ to be a low energy scale to make a sharp difference between PVLAS and stellar
production of ALPS, $Q_f$ must have a very small value. Fortunately such small charges arise
naturally in paraphoton models.

The original literature on paraphotons \cite{para} is very clear and we can be completed with
recent work around our model \cite{Mass06a,Mass06b} so there is no need of reviewing it here.

\section{A model with two paraphotons}

Let us suggest that in addition to the photon part of the standard QED lagrangian
\begin{equation}
{\cal L}_0 =-\frac{1}{4}¼ F_0^{\mu\nu}F_{0\mu\nu}+ e\ j_{0\mu} A_0^\mu
\end{equation}
we have a completely new hidden sector with active physics even at the low energy scale of PVLAS.
In this sector we find the ALP $\phi$, the fermion $f$ and two paraphotons $(A_1$ and $A_2)$, which
are the gauge boson of a gauge abelian symmetry $H$$=$$U(1)_1$$\times$$U(1)_2$. In our model, the
gauge couplings of the abelian subgroups of $H$ are equal and for simplicity are chosen to be
$e_1$=$e_2$=$e_0$=$e$, $\phi$ is completely neutral and $f$ has opposite paracharges
$Q_f^1$=$-Q_f^2$=$1$.

Our idea is that the standard QED and this proposed sector are \textit{only coupled} by the high
energy completion of the theory in which there exist ultra-heavy particles with both electric and
paraelectric charges. This particles in loops will induce kinetic mixing between the gauge bosons
of $H$ and the photon, $A_0$ gauge boson of $U(1)_0$. We also want to impose that there is a
symmetry $1\leftrightarrow 2$ in this high energy content. Then the mixing $A_0$-$A_1$ will be
equal to the mixing $A_0$-$A_2$
\begin{equation}
{\cal L}_{mix}= -\frac{\epsilon}{2} F_0^{\mu\nu}(F_{ 1\mu\nu}+F_{ 2\mu\nu}) \ \ \ . \label{lag_mix}
\end{equation}

We see then that $f$ would be completely decoupled from $A_0$ since it is coupled to the orthogonal
combination of paraphotons $A^\mu_1$$-$$A^\mu_2$ by means of our paracharge assignments. We need
then one essential ingredient in our model, mass for the paraphotons to switch  completely to the
right picture and give an electric charge to $f$.

The idea is inspired on neutrino oscillations: the state $A_+$$\propto$$A_1$$+$$A_2$ can oscillate
to $A_-$$\propto$$A_1$$-$$A_2$ if they are not propagation eigenstates. Accordingly, we choose a
maximal mixing situation $m_1=\mu\neq 0$ and $m_2=0$. In this way not only we get an electric
charge for $f$ but we make it somehow dynamical, i.e. it is going to depend on the characteristics
of the oscillation. To be more concrete we expect that the electric charge of $f$ includes a $\sin
(\mu^2 L/2\omega)$ oscillation factor ($\omega$ is the energy and $L$ the length at which the
charge is probed). But note that in the Sun the typical energy $\omega$ is much larger than in
PVLAS, and while in PVLAS $L$ is macroscopic, in the Sun the photons have a range of interaction
given by its effective mass
\begin{equation}
m_\gamma^2=\omega_P^2=\frac{4\pi\alpha n_e}{m_e} \sim \rm keV
\end{equation}
($\alpha$ is the fine structure constant and $n_e,m_e$ are the electron density and mass of the
electrons).
This considerations makes us guess a naive $\mu^2/\omega\omega_P$ suppresion of the
electric charge of $f$ in the Sun environment.

A more formal look at paraphoton models shows that we can indeed trade mixing terms as
(\ref{lag_mix}) for charge assignments. We have demonstrated it for our model in two ways: by
diagonalizing the kinetic part of the lagrangian \cite{Mass06a} and studying the amplitude of
electron-$f$ elastic scattering $e f\rightarrow e f $ \cite{Mass06b}. We get to the expected result
\begin{equation}
Q_0^f=\epsilon \frac{\mu^2}{m_0^2-\mu^2} \ \ \ .
\end{equation}

In PVLAS $m_0=0$, while in the core of the Sun $m_0$$=$$\omega_P$$\sim$$0.3$ $\rm eV$, and in the
core of a HB star $\omega_P$$\sim$$2$ $\rm keV$. There can be a big suppression of the electric
charge in this environments provided $\mu \ll \omega_P$ !

We have reach our goal of decreasing the interaction (\ref{ALP_coupling}) in the stars with respect
to the cleaner PVLAS setup. Next let me examine the values of $\epsilon,v$ and $m_f$ that respect
all the bounds on our model.

\section{Evading CAST and other constraints}

In our way of evading the astrophysical bounds we have changed so much the physics that we are lead
to additional physics consequences. Some of them are harmless or even promising but the most are
unwanted and must be evaded. The details can be read in \cite{Mass06a} but let us comment briefly
the three most important constraints.

The most relevant mechanisms of stellar energy loss in our model is plasmon decay into $f$ pairs
($\gamma^*\rightarrow \overline{f} f$) and Compton production of massive paraphotons $A_1$.
Observations of HB stars in globular clusters set the bounds \cite{Davi91,Grif86,Mass06c}

\begin{equation}
\epsilon\  \frac{\mu^2}{ \rm eV^2} <   4\times 10^{-8} \label{N1}
\end{equation}

\begin{equation}
\epsilon\  \frac{\mu^2}{ \rm eV^2} < 1.2\times 10^{-7} \ \ \ .
\end{equation}
Let us consider now the CAST limit. The production of Solar ALPs depends on the nature of the ALP
itself and three possibilities have been examined.

A) $\phi$ is a fundamental particle. The Primakoff production is suppressed by $Q_o^{f4}$, so
production takes place mainly through three body plasmon decay $\gamma^*\rightarrow \bar{f}f \phi$
which is only of order $Q_0^{f2}$. As a consequence the average ALP energy is smaller than the
plasma frequency $\omega_P \sim 0.3 \rm keV$ and thus it lies below the CAST lower threshold.

B) $\phi$ is a composite $\bar{f} f$ particle confined by new strong confining forces. The final
products of plasmon decay would be a cascade of $\phi$'s and other resonances which again would not
have enough energy to be detected by CAST.

C) $\phi$ is a positronium-like bound state of $\bar{f} f$, with paraphotons providing the
necessary binding force. As the binding energy is necessarily small, ALPs are not produced in the
solar plasma.

The other important constraint comes from the fact that the massive paraphoton $A_1$ maintains its
coupling to electrically charged standard fermions even in vacuum where $m_0=0$. It then modifies
the electromagnetic interactions at distances around $\mu^{-1}$ and it is constrained by $5^{th}$
force experiments looking for deviations of the coulomb law \cite{Bart88} (see also
\cite{Mass06c}).

\begin{figure}[t]
  \includegraphics[width=9cm]{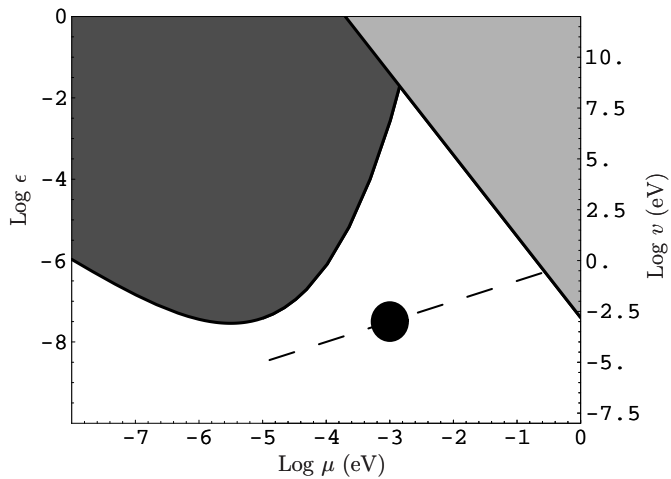}
 \caption{\it Constraints on the parameters of our model. The black
 area is excluded  by Cavendish-type experiments, and the grey area by the astrophysical
 constraint (\ref{N1}). The dashed line corresponds to $v=\mu$, and the dot to
 $v=\mu\simeq 1$ meV.
   \label{fig3} }
\end{figure}

The relevant exclusion limits in the parameter space of our model are presented in Fig.\ref{fig3}.
Notice a preferred point where all the energy scales are equal $v=\mu=1 \rm meV$ and $\epsilon$ is
very small $\sim 10^{-8}$.
It is worth to say that the introduction of the $\epsilon$-charged particle $f$ has lead also to a
reinterpretation of the PVLAS rotation in terms of $\overline{f} f$ production
\cite{Gies06,Gies06b}.

\section{Conclusions}

In the forthcoming years a bunch of optical experiments are proposed to shed some light into the
PVLAS ALP interpretation \cite{Ring06c}. If that is confirmed, which means the exciting discovery
of an axion-like particle, then to solve the PVLAS-CAST puzzle requires further new physics. We
have presented a model \cite{Mass06a} in which this physics is below the eV scale. Our model
requires some additional particles but features a high degree of symmetry. Moreover, our model has
been recently justified in some realizations of string theory \cite{Ring06d}.

\end{document}